\documentclass[fleqn,usenatbib]{mnras}

\usepackage{newtxtext,newtxmath}

\usepackage[T1]{fontenc}

\DeclareRobustCommand{\VAN}[3]{#2}
\let\VANthebibliography\thebibliography
\def\thebibliography{\DeclareRobustCommand{\VAN}[3]{##3}\VANthebibliography}

\usepackage{graphicx}	
\usepackage{amsmath}	

\def\third{{\textstyle{\frac{1}{3}}}}

\def\fthird{{\textstyle{\frac{4}{3}}}}

\def\apgt{\ {\raise-.5ex\hbox{$\buildrel>\over\sim$}}\ }
\def\aplt{\ {\raise-.5ex\hbox{$\buildrel<\over\sim$}}\ }
\def\ti{{\cal I}}
\def\Jbold{{\bf J}}
\def\Fbold{{\bf F}}

\title[Sublimation of ices in KBOs] {Sublimation of ices during the early evolution of Kuiper belt objects}

\author[Parhi and Prialnik]{Adam Parhi$^{1}$\thanks{E-mail:adamparhi@tauex.tau.ac.il}
and Dina Prialnik,$^{1}$
\thanks{E-mail:dinak@tauex.tau.ac.il}
\\
$^{1}$Department of Geosciences, Tel Aviv University, Tel Aviv, Israel
\\
}

\date{Accepted 2023 April 4. Received 2023 March 31; in original form 2023 January 11}

\pubyear{2015}

\begin{document}
\label{firstpage}
\pagerange{\pageref{firstpage}--\pageref{lastpage}}
\maketitle

\begin{abstract}
Kuiper belt objects, such as Arrokoth, the probable progenitors of short-period comets, formed and evolved at large heliocentric distances, where the ambient temperatures appear to be sufficiently low for preserving volatile ices. By detailed numerical simulations, we follow the long-term evolution of small bodies, composed of amorphous water ice, dust, and ices of other volatile species that are commonly observed in comets. The heat sources are solar radiation and the decay of short-lived radionuclides. The bodies are highly porous and gases released in the interior flow through the porous medium. The most volatile ices, CO and CH$_4$ , are found to be depleted down to the center over a time scale on the order of 100 Myr. Sublimation fronts advance from the surface inward, and when the temperature in the inner part rises sufficiently, bulk sublimation throughout the interior reduces gradually the volatile ices content until they are completely lost. All the other ices survive, which is compatible with data collected by New Horizons on Arrokoth, showing the presence of methanol, and possibly, H$_2$O, CO$_2$, NH$_3$ and C$_2$H$_6$, but no hypervolatiles. The effect of short-lived radionuclides is to increase the sublimation equilibrium temperatures and reduce volatile depletion times. We consider the effect of the bulk density, abundance ratios and heliocentric distance. At 100~au, CO is depleted, but CH$_4$ survives to present time, except for a thin outer layer. Since CO is abundantly detected in comets, we conclude that the source of highly volatile species in active comets must be gas trapped in amorphous ice.
\end{abstract}




\section{Introduction}
\label{sec:intro}
The Kuiper belt, consisting of primordial planetesimals and dwarf planets \citep{davies2008early}, is believed to be the source region for Jupiter-family comets \citep[][and references therein]{Peixinho2020}. In January 2019, almost three decades after the first detection of a small Kuiper belt object (KBO) by \cite{Jewitt1992}, the \textit{New Horizons} spacecraft visited a small, presumably pristine KBO---Arrokoth (486958)---at a mean distance of 44~au from the sun, providing close-up information that may serve to better understand comets. With an eccentricity of 0.03, Arrokoth’s orbit is nearly circular and undisturbed by the solar system distant planets. It is considered to be a member of the cold classical Kuiper belt (CCKB) population \citep{McKinnon2020}, which contains more or less dynamically undisturbed bodies, thought to be a distant relic of the solar system’s original protoplanetary disk. Therefore, Arrokoth most probably formed in place 4.6 Gyr ago and since then remained close to its current, large heliocentric distance \citep{Stern2019}.

Arrokoth was found to have a bi-lobate shape with two discrete, unequally sized lobes. The bi-lobate shape indicates that it was formed by a pair of formerly separate objects now in physical contact with one another \citep{McKinnon2020}. The merging of Arrokoth's lobes must have been gentle, since their shapes appear to be little altered although the inferred bulk density is very low \citep{Keane2022}, which means that the structure and composition of the merging objects should have been largely preserved. The size of Arrokoth, equivalent to a spherical diameter of about 19 km, is not large enough to have driven internal evolution after its formation \citep{Stern2019}. Therefore, it is expected to be a primordial planetesimal that preserves information on the physical, chemical, and accretional conditions in the outer solar nebula, and in particular, on its composition \citep{Lisse2021}.

Nevertheless, while Jupiter-family comets, descendants of KBOs, are rich in volatiles, among them hypervolatile species such as CO, CH$_4$ or N$_2$ \citep{Bocklee2017}, \textit{New Horizons} detected only methanol ice and possibly, with a high degree of uncertainty, traces of H$_2$O, CO$_2$, NH$_3$ and C$_2$H$_6$ \citep{Stern2019, Grundy2020}, all of them of moderate to low volatility. The absence of hypervolatiles at the surface was in fact expected \citep{Moore2018}. This implies that devolatilization must have taken place in Arrokoth and in KBOs in general \citep{Bockelee2001}, but the question is, to what extent. Indeed, loss of volatiles by solar radiation induced sublimation has been considered by various models and approximations \citep[e.g.][]{DeSanctis2001, Choi2002, Steckloff2021, Lisse2021, Kral2021, Davidsson2021}, but conclusions vary. For example, \cite{Kral2021} claim that KBOs larger than about 4 km can still contain CO ice after 4.6 Gyr of evolution, while \cite{Lisse2021} claim that CO and also N$_2$ and CH$_4$ not only do not exist today, but they should have been lost during the first few Myr of Arrokoth's history.

This controversy has prompted us to reconsider the question of volatile depletion in small KBOs, by detailed evolutionary calculations spanning the age of the solar system, and by checking the effect of various critical parameters. In Section~\ref{sec:analytic} we address the question by simple analytical estimates; in Section~\ref{sec:model} we briefly describe the numerical evolutionary model and the basic assumptions; in Section~\ref{sec:results} we describe and discuss the results of evolutionary calculations, and we summarize our conclusions in Section~\ref{sec:conclusions}.

\section{Analytical considerations}
\label{sec:analytic}

The survival of highly volatile ices in the interior of KBOs has been recently considered by \citet{Lisse2021} and \citet{Kral2021}, both using analytical estimates, rather than evolutionary calculations. The conclusions were widely different: taking CO as a typical example, \citet{Lisse2021} found that CO ices would be completely lost throughout objects of cometary size, including large objects such Arrokoth, while \citet{Kral2021} argued that CO should be retained to present day in objects larger than $\sim$4~km in diameter. The former estimate is based on surface energy balance, as if the entire CO ice sublimates from the surface (or at surface conditions); the latter estimate is based on time scales arguments, but ignores the latent heat of sublimation as heat sink. A different estimate is given below, to indicate what we should expect from the detailed evolutionary calculations that follow.

Consider a spherical body of radius $R$ and average density $\rho$ at a heliocentric distance $d_H$, corresponding to the Kuiper belt, say, $d_H=45$~au. Surface energy balance assuming a black body yields
\begin{equation}
    \frac{(1-A)L_\odot}{16\pi d_H^2}=\epsilon\sigma T_s^4 + F_s,
\label{eq:balance}    
\end{equation}
where $A$ is the albedo, 0.06 for Arrokoth \citep{Hofgartner2021}, $L_\odot$ is the solar luminosity, $\epsilon$ is the emissivity, $\epsilon\approx 1$, $\sigma$ - the Stefan-Boltzmann constant, $T_s$ - the surface temperature and $F_s$ - the thermal flux conducted to the interior, $F_s=K(dT/dz)_{z=0}$, $z$ being the depth and $K$, the thermal conductivity. An equilibrium temperature may be defined by 
\begin{equation}
   \sigma T_{eq}^4=\frac{(1-A)L_\odot}{16\pi d_H^2},
   \label{eq:teq}
\end{equation}
and since $F_s$ is much smaller than the other terms in equation (\ref{eq:balance}), we may assume $T_s\approx T_{eq}$. It is the flux conducted to the interior that provides the energy required for sublimation of ices (ignoring radiogenic heat sources). Bulk energy balance thus requires
\begin{equation}
    4\pi R^2 \tau F_s= \fthird\pi R^3\rho X_{\rm CO} H_{\rm CO},
\label{eq:balance1}    
\end{equation}
where $\tau$ is the period of time required for loss of CO, $X_{\rm CO}$ is the mass fraction of CO ice that is sublimated and $H_{\rm CO}$ is the latent heat of CO sublimation. The inward heat wave will raise the temperature up to the sublimation temperature of CO ice (20-25~K); therefore $F_s$ may be estimated by 
\begin{equation}
    F_s\approx K\frac{T_{eq}-T_{\rm CO}}{\ell},
\label{eq:flux}    
\end{equation}
where $\ell$ is the heat diffusion length scale corresponding to the diffusion time $\tau$,
\begin{equation}
    \ell^2\approx\frac{K\tau}{\rho c},
\label{eq:length}    
\end{equation}
with $c$ - the heat capacity. Substituting equations (\ref{eq:flux}) and (\ref{eq:length}) into equation (\ref{eq:balance1}) and rearranging terms, we obtain
\begin{equation}
    \sqrt{K\rho c} \sqrt{\tau} (T_{eq}-T_{CO}) \approx \third R \rho X_{\rm CO}H_{\rm CO},
    \label{eq:balance2}
\end{equation}
where $\sqrt{K\rho c}$ is the thermal inertia $\ti$, which has been derived from observations for several comets \citep{Groussin2013, Davidsson2013, Marshall2018} and found to range between 40-200~J~K$^{-1}$~m$^{-2}$~s$^{-1/2}$ (see Table~\ref{tab:thermal-inertia}). The equilibrium temperature at $d_H=45$~au is $T_{eq}=42$~K, therefore $(T_{eq}-T_{CO})\approx 20$~K. The latent heat of CO sublimation is $2.27\times 10^5$~J~kg$^{-1}$ \citep{Prialnik2004}. We may assume a typical cometary nucleus density, $\rho\approx 500$~kg~m$^{-3}$. Considering sublimation up to the present, we take $\tau\approx1.5\times 10^{17}$~s. Since the thermal inertia is poorly determined, we introduce a dimensionless parameter $f$, such that $\ti=f\ti_0$, where $\ti_0=40$~J~K$^{-1}$~m$^{-2}$~s$^{-1/2}$, the lowest value, according to observations.  Substituting the values of constants into equation (\ref{eq:balance2}), we obtain a relation between the three parameters, $R_{km}$ ($R$ in km), $X_{CO}$ and $f$, of the form $8.2f=R_{km}X_{CO}$, meaning that for
\begin{equation}
    R_{km}< \frac{8.2f}{X_{\rm CO}}
\label{eq:llimit}
\end{equation}
a KBO will lose its CO ice completely. Even assuming the CO abundance to be 10\% relative to water ice, which is typical of Oort cloud comets, but much larger than the CO abundance deduced for Jupiter-family comets that originate in the Kuiper belt \citep{DelloRusso2016}, and thus adopting $X_{\rm CO}\approx 0.05$, the upper limit is larger than R=150~km for $f=1$, which means that practically all KBOs should lose their CO ice entirely, and similarly, other highly volatile ices as well (in line with the conclusion of \citet{Lisse2021}). \citet{Kral2021} adopt an extreme and unlikely correction factor of 0.001 for the conductivity due to porosity, which corresponds to $f\approx 0.03$. With this value, that is, $\ti\approx 1$~J~K$^{-1}$~m$^{-2}$~s$^{-1/2}$, the upper limit for complete loss of CO is reduced to $R\sim$5~km, and in this case, the Kuiper belt may indeed be continually filled with CO produced by larger objects to present day.

\begin{table}
	\centering
	\caption{Thermal inertia of comets derived from observations.}
	\label{tab:thermal-inertia}
	\begin{tabular}{lcl} 
		\hline
		Comet & $\ti$~[J~K$^{-1}$~m$^{-2}$~s$^{-1/2}$]  &  Reference\\
		\hline
		9P/Tempel 1 & 50 - 200 & \citet{Davidsson2013} \\
		9P/Tempel 1 & < 45 & \citet{Groussin2013} \\
	    103P/Hartley 2 & < 250 & \citet{Groussin2013} \\
	    67P/C-G  & 40 -160 & \citet{Marshall2018} \\
		\hline
	\end{tabular}
\end{table}
Although the present analytic approach takes into account factors that have been ignored in the other estimates, such as the latent heat of sublimation and the fact that most of the solar energy is re-emitted as thermal radiation, it is still not entirely reliable, mainly because it assumes that sublimation occurs by an advancing front from the surface to the center. In reality, once the front recedes from the surface, the sublimated gas at the front flows both outwards and inwards. When the porous interior becomes filled with gas, the result is an isothermal structure at the sublimation equilibrium temperature. Thus sublimation takes place throughout the interior, down from the front to the center. This has been shown in a different context by detailed evolutionary calculations \citep{Prialnik2009}, and has prompted us to carry out such calculations for KBOs as well. 

\section{Evolution model and assumptions}
\label{sec:model}

The evolution model used in this study is described in detail in \citet{Prialnik1992, Prialnik2004}.
We assume a composition of dust and $N$ ice species, H$_2$O being the most abundant among them, and a highly porous structure.
The mass fractions of the volatile species are denoted by $X_{\rm s,n}$ for the solid phase and $X_{\rm v,n}$ for the vapor phase ($1\le n\le N$), and $X_d$ is the mass fraction of dust. The corresponding partial densities are by definition, the mass fraction times the density. Thus, 
\begin{equation}
\rho=\sum_n(\rho_{s,n}+\rho_{v,n}) + \rho_d
\label{eq:density}
\end{equation}
where $n$ runs over all the volatile species. The porosity is given by
\begin{equation}
\psi=1-\sum_n\rho_{s,n}/\varrho_n-\rho_d/\varrho_{d},
\label{eq:porosity}
\end{equation}
where $\varrho$ denotes the characteristic density of the non-porous solid phase (for example, $\varrho_{{\rm H}_2{\rm O}}=917$~kg~m$^{-3}$).
The energy per unit volume is given by
\begin{equation}
\rho u=\sum_{n}(\rho_{s,n}u_{s,n}+\rho_{v,n}u_{v,n}) +\rho_d u_d,
\label{eq:energy}
\end{equation}
where the specific energies $u$ are functions of temperature.
Let $\Jbold_{n}$ denote gas (vapor) fluxes, $\Fbold$ -- the heat flux, $\Fbold=-K\nabla T$, and 
$q_{n}$ -- the volume rates of sublimation of volatiles, $q=SZ(T)$, where $S$ is the surface  to volume ratio characterizing the porous medium and $Z(T)=\sqrt{m/2\pi kT}[A{\rm exp}(-B/T)-P]$ is the surface sublimation rate \citep{Mekler1990}, where $m$ is the molecular mass, $A$ and $B$ are the coefficients of the saturated vapor pressure in the Clausius-Clapeyron approximation, and $P$ is the gas pressure. The gas densities in the cases considered are very low, and hence the flow is in the Knudsen regime, thus proportional to $\nabla (P/\sqrt{T})$. The evolution
of the body's structure is described by a set of conservation laws. The mass conservation equations are:
\begin{equation}
\frac{\partial\rho_{s,n}}{\partial t}=-q_n,
\label{eq:cons-ice}
\end{equation}
\begin{equation}
\frac{\partial\rho_{v,n}}{\partial t}+\nabla\cdot \Jbold_n=q_n.
\label{eq:cons-vap}
\end{equation}
The energy conservation law yields:
\begin{equation}
\frac{\partial}{\partial t}(\rho u)+\nabla\cdot(\Fbold+\sum_{n}u_{v,n}\Jbold_n)=
-\sum_n q_n H_n+\dot Q.
\label{eq:cons-en}
\end{equation}
The second term on the left-hand-side represents the transfer of heat by conduction and advection by flowing volatiles, if present, and the terms on the right-hand-side include the rates of absorption/release of latent heat by sublimation/condensation in pores, and of radiogenic heating ($\dot Q$). The implicit assumption is that all the components of the body are in local thermodynamic equilibrium and hence a
unique local temperature may be defined. The set of  non-linear time-dependent second-order partial differential equations~(\ref{eq:cons-ice})-(\ref{eq:cons-en}) is solved numerically on a sphere. 
We adopt Arrokoth as KBO prototype, that is, a sphere having the same volume.

The solution requires initial and boundary conditions.
The inner boundary conditions are vanishing fluxes of mass and heat. 
 At the surface, the gas pressures are assumed to vanish.
The surface heat flux is given by equation~(\ref{eq:balance}).

The initial homogeneous conditions and the physical parameters adopted for all models are summarized in Table~\ref{tab:parameters}.

\begin{table}
	\centering
	\caption{Initial conditions and physical parameters.}
	\label{tab:parameters}
	\begin{tabular}{ll} 
		\hline
	Parameter &  Value \\
		\hline
		KBO radius & 9.5~km  \\
		Heliocentric distance & 45~au  \\
	    Initial uniform temperature & 16~K  \\
	    Initial dust mass fraction & 0.4 \\
	    Density & 500~kg m$^{-3}$ \\
	    Porosity & 0.6 \\
	    Heat capacity of ice & 7.49T+90~J kg$^{-1}$ K$^{-1}$ \\
	    Heat capacity of dust & 130~J kg$^{-1}$ K$^{-1}$ \\
	    Thermal conductivity (at 20~K) & 0.0035~W m$^{-1}$ K$^{-1}$ \\
	    Thermal inertia (at 20~K) & 20~J m$^{-2}$ K$^{-1}$ s$^{-1/2}$ \\
		\hline
	\end{tabular}
\end{table}

\section{Results of evolutionary calculations}
\label{sec:results}

We begin by verifying the conclusions reached by the simple analytical considerations of Section~\ref{sec:analytic}, taking into account only H$_2$O ice (with amorphous ice characteristics) and CO ice. We then proceed to more complex and realistic initial compositions.

\subsection{The simple case of H2O and CO ices}
\label{ssec:wandco}

We compute the evolution of a series of models, varying systematically one of three parameters---radius, CO to H$_2$O mass ratio and thermal conductivity (through the thermal inertia factor $f$)---and keeping all the others at the values listed in Table~\ref{tab:parameters}, until the CO ice is exhausted. We include radioactive heating by short-lived radionuclides $^{26}$Al and $^{60}$Fe as well as by long-lived radioisotopes of K, Th, and U. The parameter values and the resulting time required to emit the entire CO content are listed in Table~\ref{tab:COmodels}: model 1 is the baseline model; models 2 and 3 have higher and lower conductivities, respectively; model 4 and 5 have higher and lower radii, respectively; and models 6 and 7 have higher and lower $X_{\rm CO}/X_{{\rm H}_2{\rm O}}$, respectively. As expected, a high conductivity, small radius and low initial CO content result in a rapid loss of CO, in less than 100~Myr, with the thermal conductivity being the dominant factor. Only one out of the seven models considered---corresponding to a factor of 10 reduction in thermal conductivity---retained a (large) fraction of the initial CO beyond 4.6~Gyr. These results are consistent with the analytical prediction (\ref{eq:llimit}).
\begin{table}
	\centering
	\caption{Model parameters and results for H$_2$O and CO mixtures.}
	\label{tab:COmodels}
	\begin{tabular}{ccccr} 
		\hline
	Model & $R_{km}$ &  $X_{\rm CO}/X_{{\rm H}_2{\rm O}}$ & $f$ & $\tau [10^6 {\rm yr}]$ \\
		\hline
		1 & 9.5 & 5\% & 1 & 186 \\
		2 & 9.5 & 5\% & 2 & 45\\
		3 & 9.5 & 5\% & 0.032 & $>4.6$Gyr\\
		4 & 20   & 5\% & 1 & 518 \\
		5 & 5  & 5\% & 1 & 82\\ 
		6 & 9.5 & 11\% & 1 & 377 \\
		7 & 9.5 & 2.5\% & 1 & 80\\
	\hline
	\end{tabular}
	
Note: $f$ corresponds to a conductivity correction factor $f^2$.
\end{table}

\subsection{Multiple volatile mixtures}
\label{ssec:multiple}

\begin{figure}
	\includegraphics[width=\columnwidth]{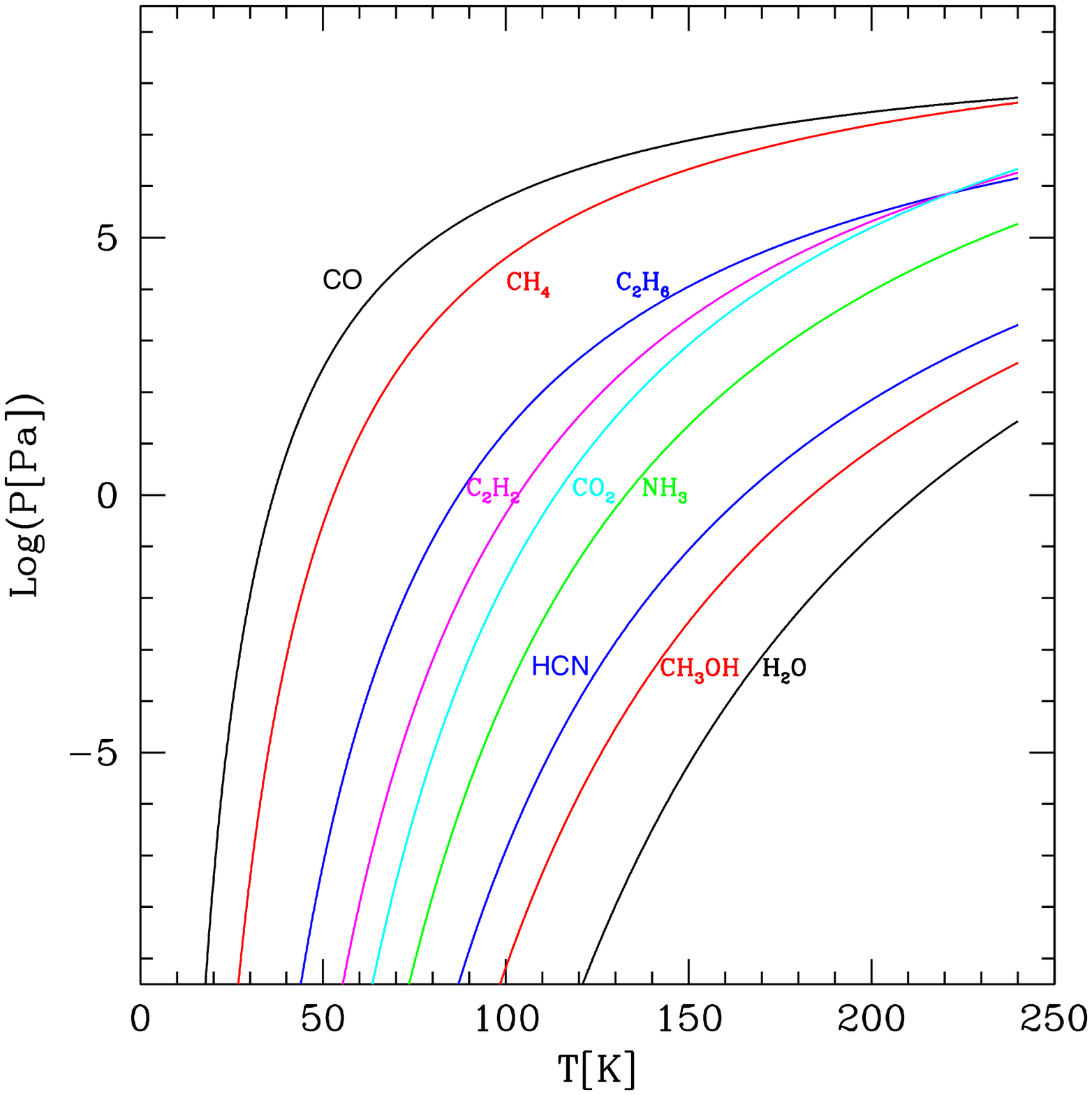}
    \caption{Saturation vapor pressures for the volatiles included in the model. Clearly, only CO, CH$_4$ will be relevant at the prevailing low temperatures, followed by C$_2$H$_6$.}
    \label{fig:clausius}
\end{figure}

We now consider a richer mixture, including dust, amorphous ice, and several other ice species detected in comets, whose saturated vapor pressures as functions of temperature are given in Fig.~\ref{fig:clausius}, of which---at the low ambient temperatures prevailing in the Kuiper belt---only CO, CH$_4$ and C$_2$H$_6$ are expected to sublimate. The question we address is, how do different species interfere with each other, competing for the same energy source.
The initial and physical characteristics are the same (as given in Table~\ref{tab:parameters}). The mass fraction of water ice is 0.51 and that of each of the other volatiles, 0.01 corresponding to 0.005~g/cm$^3$. We consider different initial abundances of the short-lived radionuclides ($^{26}$Al and $^{60}$Fe). The effect of long-lived radioactive species was found to be negligible. The calculations involving many volatile species are more time-consuming, hence the runs span 20-25~Myr, until a steady state is reached---constant production rates---and the results may be extrapolated to estimate the exhaustion time $\tau$ of the volatiles. 

\begin{table}
	\centering
	\caption{Model parameters and results for multi-ice mixtures at 20~Myr.}
	\label{tab:Mixmodels}
	\begin{tabular}{lcccc} 
		\hline
	Model & A &  B & C & D \\
		\hline
		$X_0(^{26}{\rm Al})$ & 5.37(-7)& 3.2(-8) & 0 & 0 \\
		$X_0(^{60}{\rm Fe})$ & 3.46(-7) & 8.6(-8) & 0 & 0\\
		a (au) & 45 & 45 & 45 & 100 \\
		CO depth(m) & 166 & 148 & 144 & 106\\
		CH$_4$ depth(m)& 80 & 62 & 62 & 4.4\\
		C$_2$H$_6$ depth(m)& 1.2 & 1.2 & 0 & 0 \\
		Core CO(g/cm$^3$) &6.5(-8)& 9.7(-4)& 1.0(-3) & 1.9(-3)\\
		Core CH$_4$(g/cm$^3$)&4.3(-3)& 4.1(-3)& 4.2(-3) & 5.0(-3)\\
		$Q_{\rm CO}$(molec/s) & 8.2(23)& 4.6(23) & 2.2(23)& 2.2(23)\\
		$Q_{{\rm CH}_4}$(molec/s) & 5.7(22)& 2.3(23)& 1.8(23)& 4.8(20) \\
	\hline
	\end{tabular}
 
Note: Numbers in parenthesis represent powers of 10.
	\end{table}
\begin{figure}
	\includegraphics[width=\columnwidth]{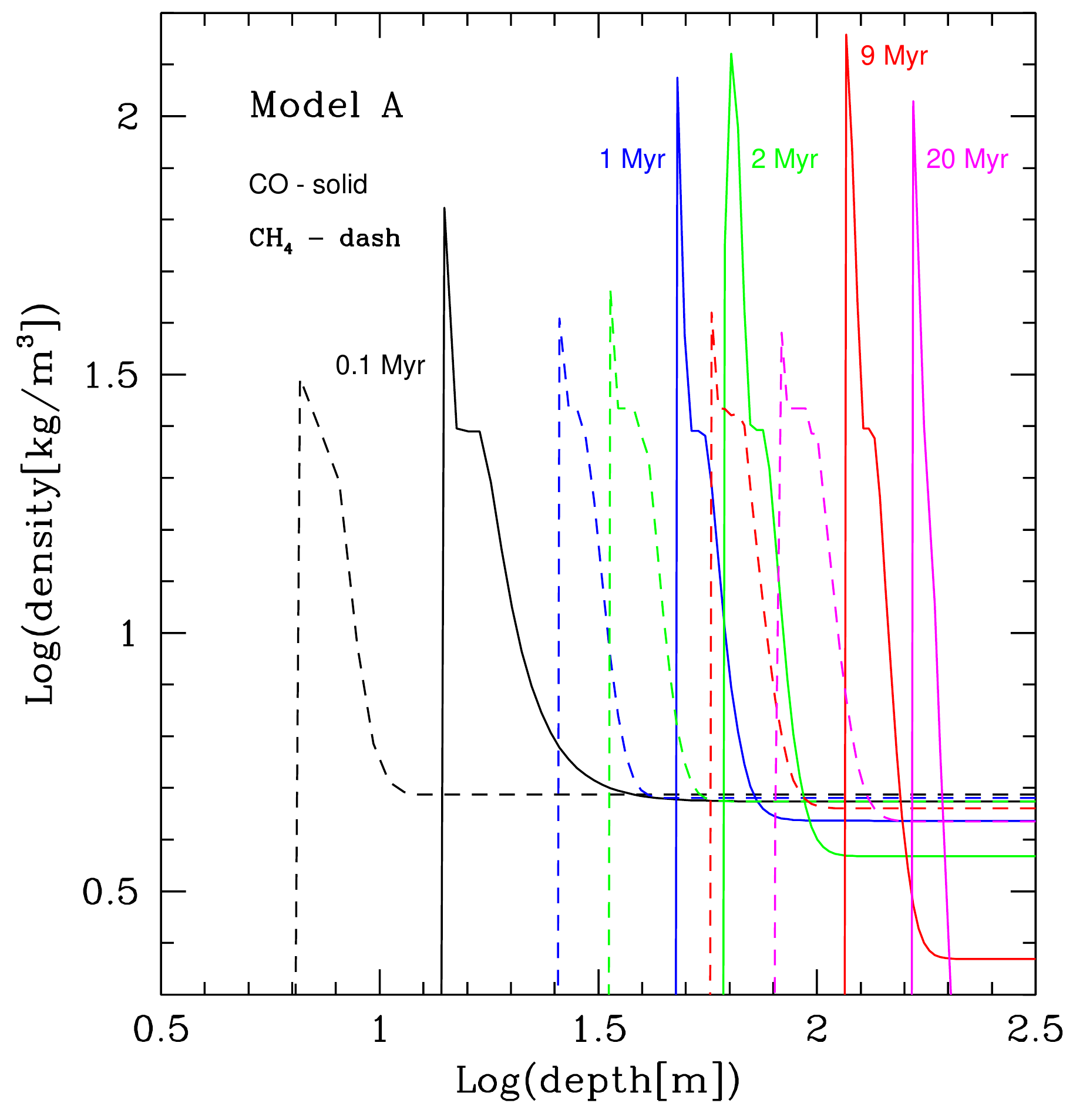}
    \caption{Advance of sublimation fronts and interior sublimation, shown by CO and CH$_4$ ice densities vs. depth at a series of times during evolution.}
    \label{fig:fronts}
\end{figure}
\begin{figure}
	\includegraphics[width=\columnwidth]{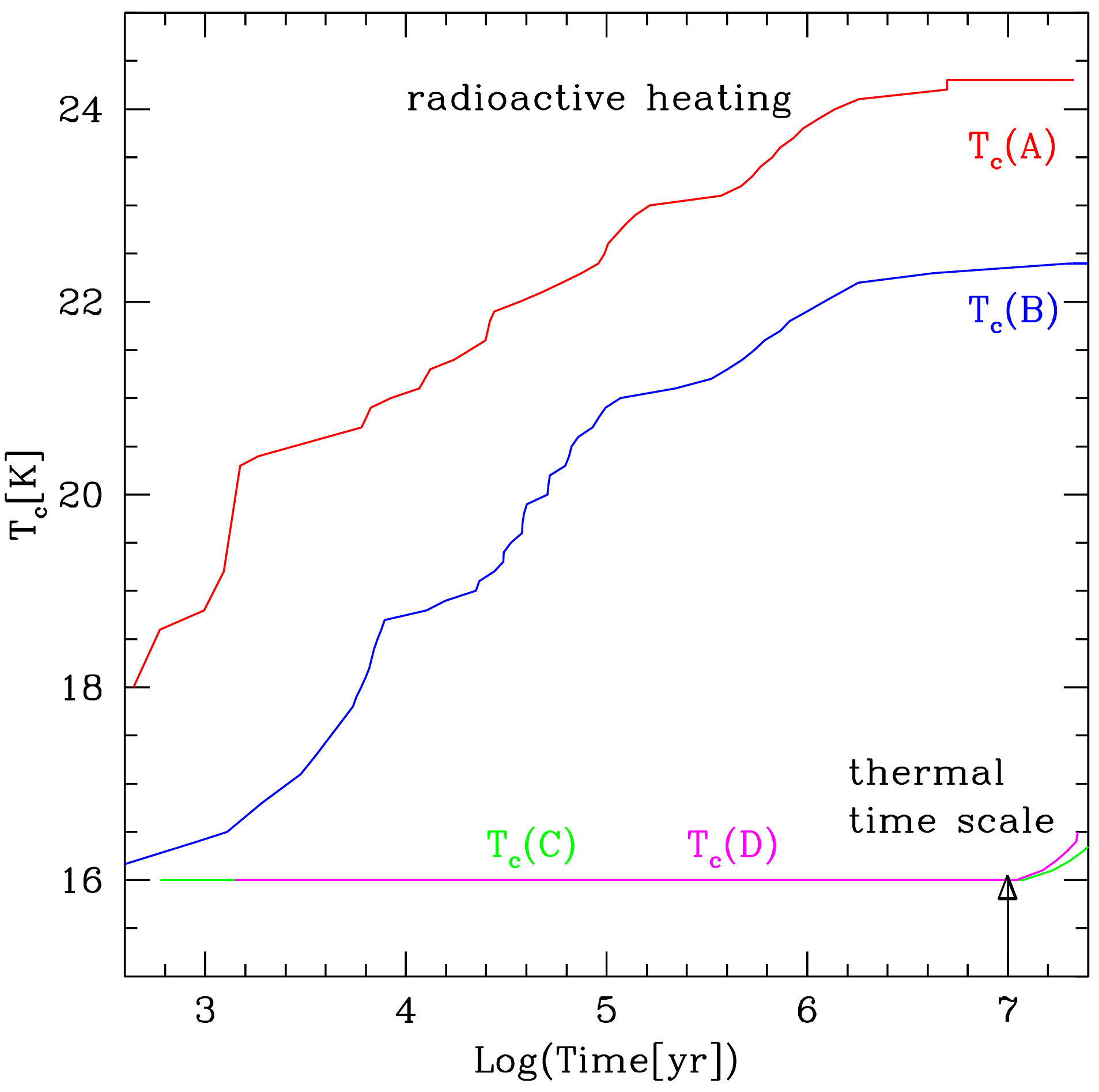}
    \caption{The evolution of the central temperature, for models A-D (as marked), as listed in Table~\ref{tab:Mixmodels}. Models A and B are heated by the decay of short-lived radioisotopes. Models C and D start to be heated by penetrating solar heat from the surface on the thermal time scale.}
    \label{fig:fluxes}
\end{figure}
The different model parameters and the main results are summarized in Table~\ref{tab:Mixmodels}. The results are given for the same evolution time of 21~Myr, when Model A has shed its entire CO content.
In Fig.~\ref{fig:fronts} we show the progress of devolatilization of CO and CH$_4$ for Model A, by the advance of sublimation fronts and by the gradual interior depletion, as illustrated by the CO and CH$_4$ density profiles at several moments in time.

The sensitivity to temperature is very high: a small temperature difference, only a few Kelvins, results in a large difference between the CO and CH$_4$ evaporation fluxes (see production rates $Q$ in Table~\ref{tab:Mixmodels}). This is easily apparent in Fig.~\ref{fig:clausius}. Therefore, in the case of Model A, CO evaporates quickly, while CH$_4$ takes a very long time to be exhausted, over 300~Myr. In Model B, where the internal temperatures are slightly lower as a result of less radioactive heating, the rates of evaporation are much closer, hence CO takes a longer time, while CH$_4$, a shorter time compared to case A, to completely evaporate. 
On average, CO is lost on timescales of a few tens of Myr, while CH$_4$ is lost later, on timescales of a few hundred Myr. Only the model for which a larger heliocentric distance (100~au) was assumed, retained a large fraction of its initial CH$_4$ for the age of the solar system. We also computed a model at an even larger distance, 220~au, and found it to retain its volatiles almost intact; after 4.6~Gyr, the CO sublimation front has receded to a depth of only 10~m. These results should be relevant for the dynamics of the Kuiper belt, since scattering to more distant orbits may stop the devolatilization process.

\subsubsection{The effect of radioactive heating}
\label{sssec:radio}

Radioactive heating of small bodies by short-lived radioisotopes has been considered in a number of studies, under various assumptions: volatiles trapped in amorphous ice, but not as free ices; gases released in the interior escaping freely to the surface; long delay times (low initial abundances of the radioactive isotopes); only CO as representative of hypervolatiles \citep[see][for comprehensive reviews]{Jewitt2007, McKinnon2008}. The present study includes the most abundant volatile species observed in comets as a mixture of ices, and follows the long-term evolution taking into account the diffusion of volatiles through the porous medium, as described in Section~\ref{sec:model} and abundances of the short-lived radioisotopes corresponding to 0 and 3~Myr accretion times (see Table~\ref{tab:Mixmodels}).

A test model was run for 4.6~Gyr at 45~au, without volatile ices, adopting the radioactive abundances of Model A; the internal temperatures reached 150~K and complete crystallization resulted except for a $\sim$meter-thick outer layer, in agreement with previous studies. By contrast, for models A and B, which included ices of CO, CH$_4$ and C$_2$H$_6$, internal temperatures barely exceeded 25~K, as shown in Fig.~\ref{fig:fluxes}. This result can be easily understood as follows. 

The highest radioactive energy production rate, corresponding to $t=0$, is 
$\dot Q_{\rm max}=\rho_d[X_0(^{26}{\rm Al})H_{\rm Al}/\tau_{\rm Al}+X_0(^{60}{\rm Fe})H_{\rm Fe}/\tau_{\rm Fe}]$, where $\rho_d$ is the dust mass per unit volume, $H$ is the energy released per unit mass of a radioactive isotope, $X_0$ is the isotope's abundance relative to dust, and $\tau$ is the radioactive decay time scale. The rate of energy absorption due to 
volatile sublimation per unit volume is $q=\sum q_n H_n$ (see Eq.~\ref{eq:cons-en}), with
\begin{equation}
q_n=S\sqrt{\frac{m_n}{2\pi kT}}[A_n\exp{(-B_n/T)}-P_n]. 
\label{eq:surate}
\end{equation}
In thermodynamic equilibrium, the gas pressure equals the saturated vapor pressure; sublimation in the interior occurs very close to equilibrium \citep[see, e.g.][]{Coradini1997}, such that $A_n\exp{(-B_n/T)}-P_n=\varepsilon A_n\exp{(-B_n/T)}$, with $\varepsilon\ll 1$. When thermal equilibrium is achieved, where the rate of radioactive energy release equals the rate of energy absorbed in sublimation,
\begin{equation}
  \dot Q(t) = S\sum_n H_n\sqrt{\frac{m_n}{2\pi kT}}\varepsilon A_n\exp{(-B_n/T)},
\end{equation}
it yields $T\aplt{25}$~K for the parameter values assumed (where the exact value of $\varepsilon$ has a minor effect). It so happens that the decay time of the short-lived radioisotopes is of the same order of magnitude as the thermal diffusion time scale of the solar energy absorbed at the surface. Therefore, when the radioactive energy source declines, solar energy replaces it (see Fig.~\ref{fig:fluxes}). The surplus radioactive energy is radiated away.

\subsection{Cometary composition}
\label{ssec:cometN2}

We have seen that mixture of ices of similar volatility lead to widely different timescales for the depletion of these ices. It is thus of interest to adopt initial compositions that are typical of comets \citep{Bocklee2017}, rather than arbitrarily equal abundances that are only of academic interest by helping to understand the process. 
\begin{table}
	\centering
	\caption{Av and Bo models initial conditions and physical parameters.}
	\label{tab:av-bo-parameters}
	\begin{tabular}{lll} 
		\hline
	Parameter &  Av & Bo \\
		\hline
		KBO radius[~km] & 4.75 & 4.75  \\
		Heliocentric distance[~au] & 20 & 20  \\
	    Density[~kg m$^{-3}$] & 500 & 500 \\
	    CO & 5.8(-2) & 2.35(-2) \\
	    CH$_4$ & 4.05(-3) & 5(-3) \\
	    C$_2$H$_6$ & 5.35(-3) & 1.6(-3) \\
	    NH$_3$ & 2.5(-3) & 1.75(-3) \\
	    C$_2$H$_2$ & 1.35(-3) & 8(-4) \\
        N$_2$  &  - & 5(-4)\\ 
        CO$_2$  &  8.12(-2) & -\\ 
		\hline
	\end{tabular}
 
Note: Numbers in parenthesis represent powers of 10.
\end{table}
\begin{figure}
	\includegraphics[width=\columnwidth]
{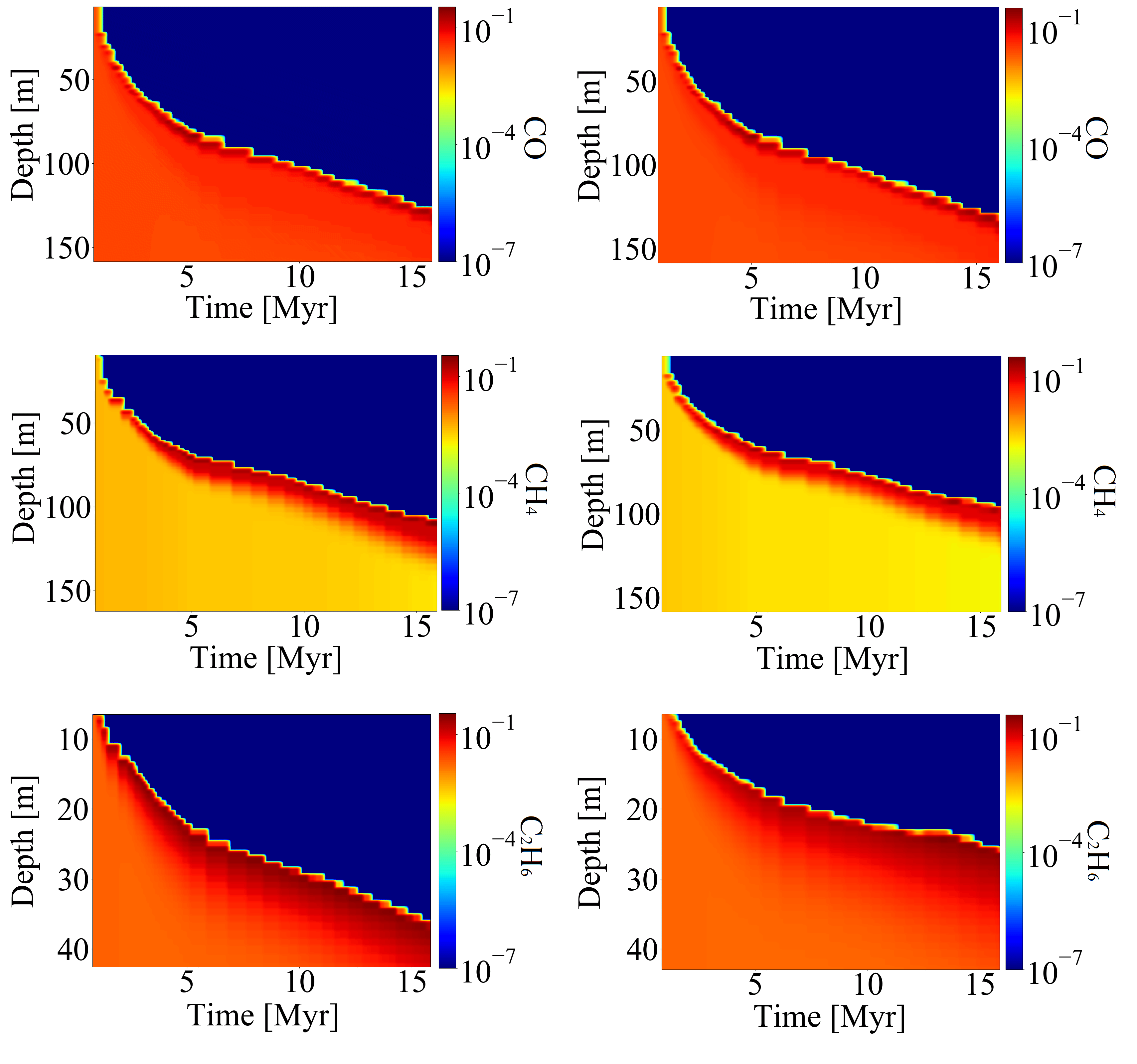}
    \caption{Depth profiles of CO, CH$_4$, and C$_2$H$_6$ for model Bo with N$_2$ (left panels) and model Av (right panels) at a heliocentric distance of 20~au. The profiles for model Bo without the N$_2$ (not shown) were found to be very similar. The volatile enhanced layer ahead of the sublimation front is clearly seen.
    The roughness of the profile is only due to sparse sampling of the results.}
    \label{fig:depth_profile_bo_volatiles}
\end{figure}

We thus consider two comet models, one with typical average amounts of volatiles---model Av---and one with the composition of comet 9P/Borrely \citep{boice2002deep}---model Bo. The initial conditions and physical parameters are given in Table~\ref{tab:av-bo-parameters}. We also consider a wider range of distances as potential locations for the early evolution of comets: from 20~au---representing the primordial disc of planetesimals---through 40~au---representing the CCKB---to 80~au---representing the outer edge of the Kuiper belt \citep[see][and references therein]{Davidsson2021}. 

In Fig.\ref{fig:depth_profile_bo_volatiles} we compare the behaviour of the most volatile species---CO, CH$_4$ and C$_2$H$_6$---by looking at the depths of the corresponding sublimation fronts as a function of time, for both models, Bo and Av. For model Av, we note that the depletion depths follow the order of volatility, with the deepest sublimation front corresponding to the most volatile species (CO), despite the fact that the initial abundances follow the reverse order, CO being the most abundant. For model Bo, the very low initial abundance of C$_2$H$_6$ comes into play, with a deeper sublimation front than that of CH$_4$. 

\begin{figure}
	\includegraphics[width=\columnwidth]{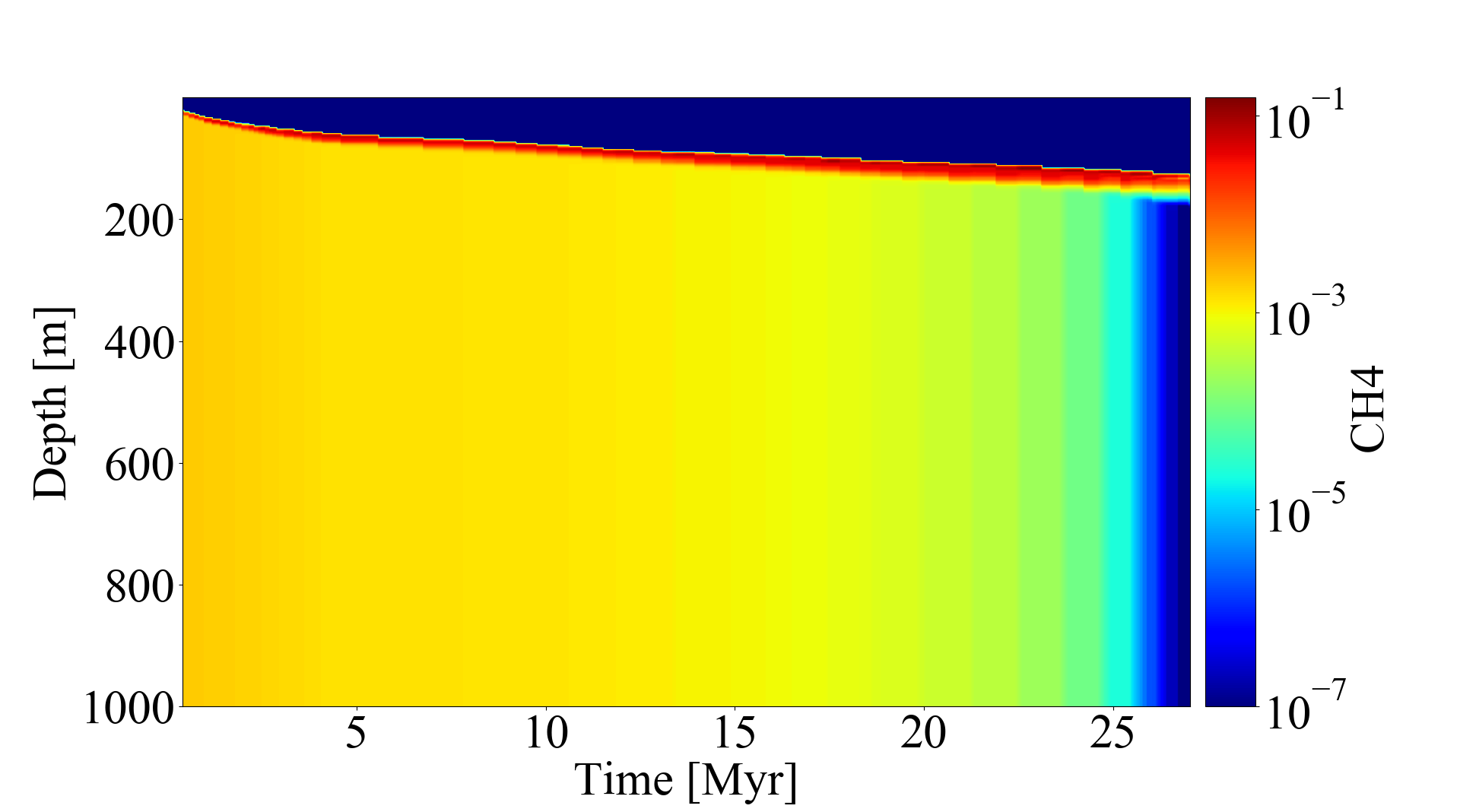}
    \caption{Depth profile of CH$_4$ for model Av after $\sim$27~Myr of evolution. The advance of the sublimation front, the build-up of a dense layer behind the front and the gradual internal depletion are clearly seen. The abundances are the same from the bottom depth of 1~km down to the center.}
    \label{fig:Av-CH4}
\end{figure}

The less volatile species---C$_2$H$_2$, NH$_3$ and CO$_2$---sublimate as well, but much more slowly. In Table~\ref{tab:fronts} we show the sublimation front depths for all the molecular species taken into account in model Av after 27~Myr. Here we note in particular the effect mentioned in Section~\ref{sec:analytic} and illustrated in Fig.~\ref{fig:Av-CH4}, namely, the gradual depletion of CH$_4$ throughout the interior, in addition to the inward advance of the sublimation front from the surface. In fact, the former process is much more efficient than the latter.
The emerging general conclusion is that relative bulk abundance ratios are not conserved during evolution.
\begin{table}
\centering
	\caption{Sublimation front depths for the Av model after 27~Myr.}
	\label{tab:fronts}
\begin{tabular}{l|cccccc}
\hline
Molecule     & CO & CH$_4$ & C$_2$H$_6$ & C$_2$H$_2$ & CO$_2$ &  NH$_3$ \\ 
Depth{[}m{]} &  176 & 132$^*$ & 37  & 19  & 0.55 & 0.10  \\ \hline
\end{tabular}

$^*$ CH$_4$ is depleted throughout, except for a 25~m-thick layer beneath the front.
\end{table}

The N${_2}$ molecule has been rarely observed in comets as compared with the ubiquitous hypervolatile species, CO, CH$_4$ and C$_2$H$_6$ (see \cite[e.g.,][]{Opitom2019}). It was detected in comet 67P/Churyumov-Gerasimenko by the ROSINA instrument, at a very low abundance relative to CO ($5.7\times 10^{-3}$, \cite{Rubin2015}); the highest N$_2$/CO ratio, reported by \cite{Cochran2018}, being 0.15 for comet C/2016 R2 (PanSTARRS). We tested the effect of adding N$_2$ to our inventory of volatiles and found that, while it was the first to be depleted, it did not affect the temperature evolution, nor the sublimation of the other volatile species to an appreciable extent.

\subsubsection{The effect of heliocentric distance}
\label{sssec:distance}

\begin{figure}
	\includegraphics[width=\columnwidth]{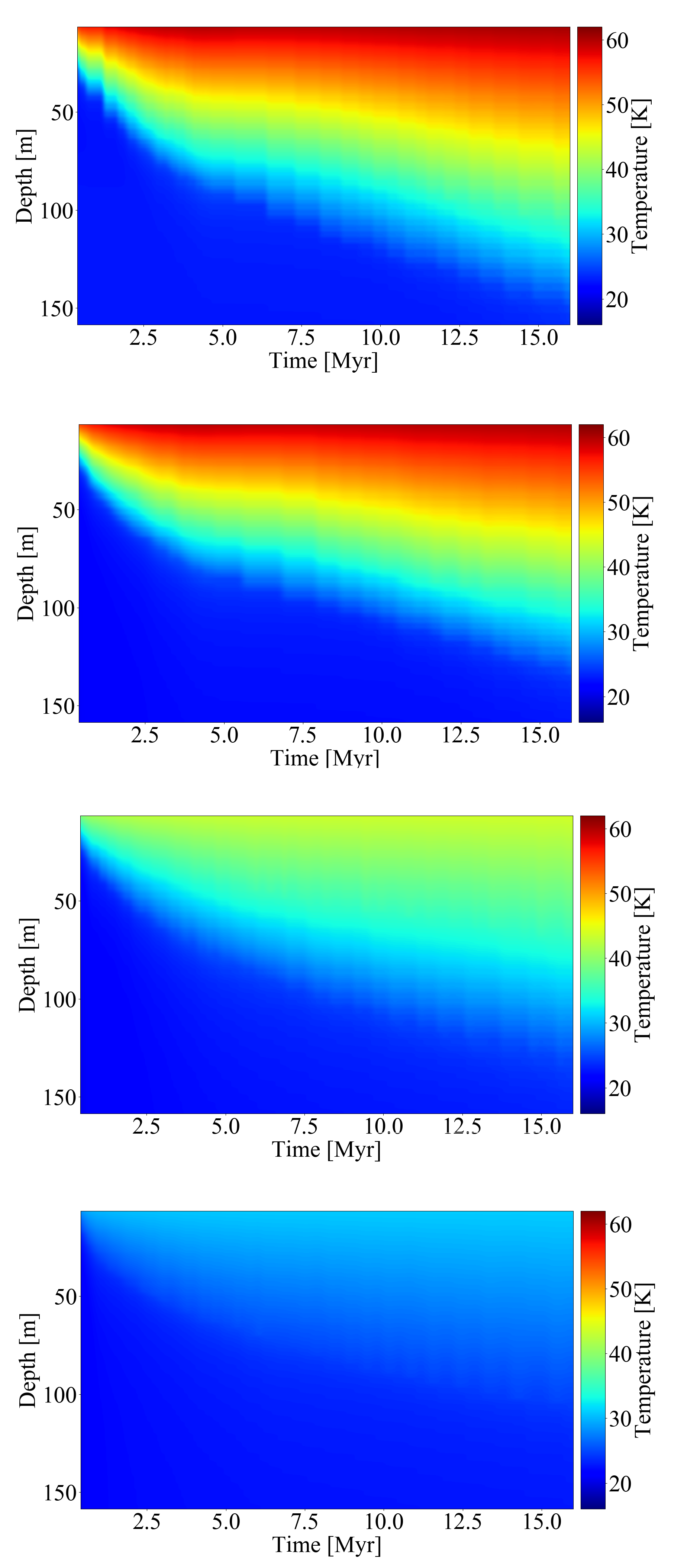}
    \caption{Evolution of the temperature profile during 15~Myr, for model Bo at a heliocentric distance of 20~au (top panel) and for model Av at 20~au, 40~au and 80~au (next panels from top to bottom). The differences between models Bo and Av---different compositions---at the same distance are barely noticeable. }
    \label{fig:volatile_production_rates_by_dst}
\end{figure}
We first show in Fig.~\ref{fig:volatile_production_rates_by_dst} the evolution of the temperature profile along 15~Myr, at the heliocentric distances considered. The effect of the diminishing energy source is obvious and expected. The dependence of volatile depletion on heliocentric distance is more complicated.

\begin{figure}
	\includegraphics[width=\columnwidth]{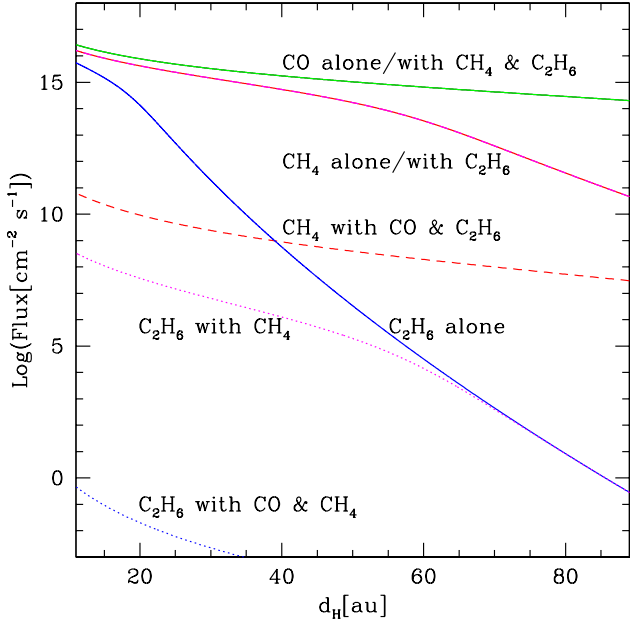}
    \caption{Production rates of volatiles, by themselves (marked \textit{alone}) or in mixtures (marked \textit{with}), as function of heliocentric distance, obtained from the solution of Eq.(\ref{eq:fluxes}).}
    \label{fig:comflux}
\end{figure}

A mixture of ices of comparable volatility, competing for a given heat source, behaves very differently from each ice by itself, given the same heat source. Moreover, the relative production rates vary considerably with the power of the energy source, namely with heliocentric distance, when the source is solar energy. We illustrate the effect by a simple example: assuming that the absorbed solar energy is consumed by free sublimation in the outer layers, we solve the equation
\begin{equation}
    \frac{L_\odot}{16\pi d_H^2}-\sigma T^4 = \sum_n H_n\sqrt{\frac{m_n}{2\pi kT}}A_n\exp{(-B_n/T)},
    \label{eq:fluxes}
\end{equation}
for $T(d_H)$ and derive the molecular fluxes. We consider two mixtures: (a) CO, CH$_4$ and C$_2$H$_6$, and (b) CH$_4$ and C$_2$H$_6$, and for comparison, each species by itself. The results for the corresponding fluxes as function of $d_H$are shown in Fig.~\ref{fig:comflux}.
We note that the most volatile species (CO, when present, or CH$_4$ in the presence of C$_2$H$_6$) are dominant, and that the flux ratios depend strongly on distance.

The length of time required for a complete loss of volatiles, as derived for the composition of model Av, assuming different heliocentric distances, is summarized in Table~\ref{tab:emission}. Due to the large abundance difference between CO and CH$_4$, it is difficult to give a more accurate estimate for the depletion time of the latter; its abundance declines faster than that of the more abundant CO, but there are still low traces left while CO is exhausted. It is interesting to note that C$_2$H$_6$ survives to present time at distances beyond $\sim 40$~au.
\begin{table}
	\centering
	\caption{Volatile depletion times [Myr] obtained for model Av at different heliocentric distances.}
	\label{tab:emission}
	\begin{tabular}{cccc} 
		\hline
	Distance [au]  &  CO & CH$_4$ & C$_2$H$_6$ \\
		\hline
		20 &  39 & $\sim 20$ & 680\\
		40 &  48 & $\sim 20$ & $>4.6$Gyr\\
		80 &  84 & 3.6 Gyr & $>4.6$Gyr\\
	\hline
	\end{tabular}
\end{table}

\subsubsection{The effect of bulk density}
\label{sssec:parameters}

The comet nucleus density is bound to affect the volatile depletion time scales because the amount of volatiles increases with bulk density (for given abundance ratios), while the heat source remains the same.
Arrokoth’s bulk density must be at least 290~kg~m$^{-3}$ \citep{ spencer2020geology}. The most accurate comet nucleus density, obtained for 67P/Churyumov-Gerasimenko, is 533~kg~m$^{-3}$ \citep{Patzold2016}. Higher densities have been derived for CCKB objects \citep{Mueller2020}. We therefore repeat the evolutionary calculations for model Av at 20~au, with two additional bulk densities, 300 and 700~kg m{$^{-3}$}, besides the baseline 500~kg m{$^{-3}$}. 

In Table~\ref{tab:densities} we compare the lengths of time required for the sublimation fronts of the three hypervolatiles to reach the depth of 55~m for each bulk
density. This depth is arbitrarily chosen for illustration purposes; it is the heat diffusion depth corresponding to a full orbital period.
The correlation between density and sublimation time is not straightforward. A higher density, which means a lower porosity, increases the thermal conductivity, which results in higher subsurface temperatures. Therefore, sublimation will start earlier and proceed at a higher rate (lesser sublimation time). This effect is limited to the heat diffusion depth. At higher depths, the temperature will be far less affected, but then, the fact that higher densities mean more volatile ices, leads to longer depletion times.
As expected, the times increase with increasing density (i.e., absolute volatile abundances), but the trend is far from linear and, moreover, differs among species. Also as expected, the times increase with decreasing volatility. This is due to the fact that the temperature is determined by the phase equilibrium of the most volatile among the gases, in this case CO (see also Fig.~\ref{fig:comflux}). Thus CO consumes the largest fraction of the solar energy. When CO is exhausted, the temperature is determined by the next volatile, hence the behaviour depends on several factors and the result is less intuitive. 

\begin{table}
	\centering
	\caption{Recess times of the sublimation front [Myr] to a depth of 55~m for the composition of model Av at 20~au, for different bulk densities.}
	\label{tab:densities}
\begin{tabular}{cccc}
\hline
Density [kg m$^{-3}$] & CO   & CH$_4$ & C$_2$H$_6$       \\ \hline
                        300                 & 1.62 & 3.38   & 39.23           \\
 500                 & 1.7 & 3.45   & >100            \\
                        700                 & 2.79 & >4.5  & >150 \\ \hline
\end{tabular}
\end{table}

\section{Summary and Conclusions}
\label{sec:conclusions}

Kuiper belt objects, such as Arrokoth, which was visited by the \textit{New Horizons} space mission in 2019, have probably formed and evolved at large heliocentric distances, where the ambient temperatures were sufficiently low for preserving volatile ices, if not at the surface, then perhaps in the interior. The purpose of the present study was to follow the long-term evolution of small bodies, composed of amorphous water ice, dust grains and ices of other volatile species that are commonly observed in comets: CO, CH$_4$, C$_2$H$_6$, CO$_2$, NH$_3$, C$_2$H$_2$, N$_2$, etc. The heat sources were solar radiation and the decay of short-lived radionuclides, $^{26}$Al and $^{60}$Fe. We adopted a high ice to dust ratio, to obtain upper limits for depletion time scales (worst scenario). We considered highly porous bodies, so that gases released in the interior could flow through the porous medium. 

Under the assumed conditions---a circular orbit at distance of 44~AU, a fast-rotating body (namely, a uniform surface temperature), and a density of 500~kg m$^{-3}$---the most volatile ices, CO and CH$_4$, were found to be depleted even at the center of the body over a time scale on the order of 100~Myr. Sublimation fronts advanced from the surface inward, and when the temperature in the inner part rose sufficiently, bulk sublimation throughout the interior reduced gradually the volatile ices content until these ices were completely lost. All the other ices survived. 

The analysis of Arrokoth's observations lends some support to our results. Although only methanol was detected at a high degree of confidence, all these ices (H$_2$O, CO$_2$, NH$_3$ and C$_2$H$_6$) were invoked to explain the data---albeit with very low probability \citep{Grundy2020}---but none of the hypervolatiles was found compatible with it. 

It is interesting to compare the results of our evolutionary calculations with those obtained in the past by means of similar numerical models. \cite{Choi2002} addressed the same problem, but they assumed that any gas released in the interior escapes instantaneously, even if sublimation occurs at large depths. This assumption has two effects: it shortens the volatile depletion time and allows the internal temperature to rise significantly, since it is not controlled by the pressure equilibrium attained in pores. \cite{DeSanctis2001} did consider the diffusion of gas sublimated in the interior and their results are very similar to ours. However, due to the limited computational means of two decades ago, they considered only a few cases and only CO as representative of hypervolatile species.

Our main conclusions may be summarized as follows:
\begin{itemize}
    \item {} At the Kuiper belt distance, it is safe to assume that objects of cometary size (even up to tens of km in radius) have lost their most volatile ices---if those were included in the initial composition as ice mixtures (rather than trapped in water or CO$_2$ ices)---on time scales that are much shorter than the age of the solar system.
    \item {} Volatile mixtures that sublimate concomitantly do not retain the initial mass ratios. Therefore, if for some dynamical reason, the sublimation process is stopped (e.g., by migration to more distant regions), the chemical composition of an object will not reflect the ambient composition in any region of the solar nebula.
    \item{} Sublimation times are affected by the density of the body in two opposing ways: a higher density (lower porosity) increases thermal conductivity, which results in higher internal temperatures; therefore, volatile loss is enhanced. However, a higher density also implies a larger ice mass for the same solar energy input, leading to longer volatile exhaustion times. 
    \item {} Volatile depletion evolves both by a receding sublimation front and by internal bulk sublimation in pores. Gas flows outwards from the interior and in both directions at the sublimation front. As a result, a denser layer forms behind the front, enriched in ice of the active species. 
    \item{} The effect of long-lived radionuclides is negligible. The energy released by short-lived radionuclides reaches thermal equilibrium with sublimation at temperatures typical of the hypervolatiles sublimation rates (around 20~K).
    \item{} At a distance of 100~au, CO is depleted, but CH$_4$ survives to present time, except for a few meters thick outer layer. At a distance of 200~au, even CO survives at a depth of $\sim$10~m. 
\end{itemize}

\noindent Since KBOs are considered to be the progenitors of short-period comets, and since CO is abundantly detected in cometary comae, the conclusion of this study is that the source of highly volatile species in active comets is gas trapped in amorphous ice, and possibly in CO$_2$ ice. This conclusion gains support from the observational results of e.g., \cite{Roth2020}, who find that production rates of CO, CH$_4$ and C$_2$H$_6$ in comet 21P/Giacobini-Zinner correlate with that of H$_2$O. In fact, the primordial presence of hypervolatile ices in KBOs protects the amorphous ice and the volatiles trapped in it, by keeping the temperature far below the water ice crystallization range. Thus although as independednt species, hypervolatiles are most probably lost before comets become active, they ensure that their trapped counterparts will be detected in cometary activity, escaping from crystallizing amorphous ice.


\section*{Data Availability}

The derived data generated in this research will be shared on reasonable request to the corresponding author.



\bibliographystyle{mnras}
\bibliography{kbos} 








\bsp	
\label{lastpage}
\end{document}